\documentclass[conference]{IEEEtran}
\IEEEoverridecommandlockouts
\usepackage[utf8]{inputenc}
\usepackage{cite}
\usepackage{overpic}
\usepackage{amsmath,amssymb,amsfonts}
\usepackage{bm}
\usepackage{graphicx}
\usepackage{textcomp}
\usepackage{xcolor}
\usepackage{float}
\usepackage{amsthm}
\usepackage{graphicx}
\usepackage{epstopdf}
\usepackage{amsmath,bm,bbm}
\usepackage{amsfonts}
\usepackage{amssymb}
\usepackage{color}
\usepackage{multirow}
\usepackage{multicol}
\usepackage{soul,xcolor}
\usepackage{setspace}

\usepackage{longtable}
\usepackage{mathtools}
\usepackage{subfig}
\usepackage{tabulary}
\usepackage{epsfig}
\usepackage{caption}
\usepackage{booktabs}
\usepackage{blindtext}
\usepackage{adjustbox}
\usepackage{hyperref}
\usepackage{dirtytalk}
\usepackage{comment} 

\usepackage{tikz}
\usetikzlibrary{positioning} 
\usepackage{algorithm}

\usepackage{algpseudocode}

\definecolor{mypurple}{HTML}{9933FF}
\definecolor{mygreen}{HTML}{009900}

\theoremstyle{plain}

\newcommand{\vect}[1]{\mathbf{#1}}

\def\Ttran{\mbox{\tiny $\mathrm{T}$}}
\def\CN{\mathcal{N}_{\mathbb{C}}} 
\def\imagunit{\mathsf{j}} 
\begin{document}
\bstctlcite{IEEEexample:BSTcontrol}
\makeatletter
\newcommand*{\rom}[1]{\expandafter\@slowromancap\romannumeral #1@}
\makeatother

\title{Green One-Bit Quantized Precoding in Cell-Free Massive MIMO \vspace{-0.4cm}  }

\author{

 \IEEEauthorblockN{Salih Gümüsbuğa$^{*\ddagger}$, Ozan Alp Topal$^\dagger$, and \"Ozlem Tu\u{g}fe Demir$^*$}
\IEEEauthorblockA{ {$^*$Department of Electrical-Electronics Engineering, TOBB ETÜ, Ankara, Turkiye}
\\ 
 {$^\dagger$Department of Computer Science, KTH Royal Institute of Technology, Stockholm, Sweden} \\
 {$^\ddagger$Meteksan Savunma Sanayii A.Ş.
  Ankara, Turkiye}
\\
		{Email: s.gumusbuga@etu.edu.tr, oatopal@kth.se, ozlemtugfedemir@etu.edu.tr }
}
\vspace{-1cm}

\thanks{ The work by S. Gümüşbuğa and \"O. T. Demir was supported by 2232-B International Fellowship for Early Stage Researchers Programme funded by the Scientific and Technological Research Council of Turkiye. Additionally, S. Gümüşbuğa is supported by Meteksan Savunma Sanayii A.Ş. } 

}

\maketitle
\begin{abstract}
Cell-free massive MIMO (multiple-input multiple-output) is expected to be one of the key technologies in sixth-generation (6G) and beyond wireless communications, offering enhanced spectral efficiency for cell-edge user equipments by employing joint transmission and reception with a large number of antennas distributed throughout the region. However, high-resolution RF chains associated with these antennas significantly increase power consumption. To address this issue, the use of low-resolution analog-to-digital and digital-to-analog converters (ADCs/DACs) has emerged as a promising approach to balance power efficiency and performance in massive MIMO networks. In this work, we propose a novel quantized precoding algorithm tailored for cell-free massive MIMO systems, where the proposed method dynamically deactivates unnecessary antennas based on the structure of each symbol vector, thereby enhancing energy efficiency. Simulation results demonstrate that our algorithm outperforms existing methods such as squared-infinity norm Douglas-Rachford splitting (SQUID) and regularized zero forcing (RZF), achieving superior performance while effectively reducing power consumption. 
\end{abstract}

\vspace{-2mm}

\section{Introduction}

Massive MIMO (multiple-input multiple-output) is a cornerstone technology in modern cellular networks. By increasing the number of antennas at the base station, interference can be effectively mitigated and favorable propagation conditions are ensured. This enables the use of simple yet powerful linear precoding schemes—such as zero forcing (ZF), minimum mean squared error (MMSE), and regularized zero forcing (RZF)—to significantly enhance system capacity \cite{Larsson2014-ct}, \cite{Lu2014-ar}. While these techniques perform well under known or estimated channel conditions, they typically assume the presence of infinite-resolution digital-to-analog converters (DACs) in the RF chain. In practical systems with a moderate number of antennas, DACs with 12- to 16-bit resolution are sufficient. However, as antenna counts grow, hardware complexity and power consumption become major concerns, motivating the use of low-resolution DACs in the downlink.

In cell-free massive MIMO, a large number of distributed access points (APs), each with a few antennas, jointly serve the user equipments (UEs). Although each AP is simpler than a co-located massive MIMO base station, the total number of RF chains across the network can still be substantial. To support dense and cost-sensitive deployments, low-resolution DACs provide a compelling means to reduce hardware cost and power consumption. In recent years, quantized precoding has garnered significant attention as an energy- and cost-efficient approach that leverages low-resolution DACs, typically operating at 1 to 3 bits. In \cite{Jacobsson2017a}, the squared-infinity norm Douglas-Rachford splitting (SQUID) algorithm was proposed for multi-user massive MIMO systems and evaluated against several 1-bit quantized precoders, including ZF, semidefinite relaxation (SDR), Wiener filter (WF), and maximum ratio transmission (MRT). The quantization behavior of the WF was further analyzed in terms of uncoded bit error rate (BER) in \cite{Mezghani2009-fy}. In \cite{Yuan2020-hz}, a power-constrained 1-bit precoding scheme was introduced that allowed certain antennas to transmit zero power. A comparative analysis between 1-bit ZF and maximum likelihood precoding was provided in \cite{Saxena2016-gs}, while a low-complexity 1-bit quantized precoder was proposed in \cite{Park2019-sz} and benchmarked against quantized ZF.

Another effective strategy for reducing power consumption is antenna selection, where only a subset of antennas is activated based on transmission needs. In \cite{Tang2018-rw}, an antenna selection method for massive MIMO was proposed, and a computationally efficient algorithm was later introduced in \cite{Gao2018-te}. Additional contributions on antenna selection can be found in \cite{Joung2016-mp} and \cite{Gorokhov2003-mu}.

The joint design of quantized precoding and antenna selection has not yet been explored in the context of downlink cell-free massive MIMO systems. Building upon \cite{Jacobsson2017a}, in this study, we propose a novel quantized precoding and antenna selection algorithm specifically tailored for cell-free architectures. Our proposed algorithm builds upon the group-sparsity concept applied to one-bit quantized precoding, and is conceptually aligned with our earlier work in \cite{topal2024energy}, which investigates energy-efficient cell-free architectures with constrained fronthaul. While \cite{topal2024energy} focused on AP activation strategies and fronthaul scheduling, here we bring energy-awareness into the symbol-level precoding stage, dynamically selecting which antennas to activate on a per-symbol basis. Specifically, we cast the original non-convex optimization—jointly minimizing the mean squared error (MSE) of UE signals and the number of active antennas—into a convex relaxation using group lasso regularization. This formulation enables structured sparsity and allows the precoder to deactivate antennas with minimal signal contribution, effectively reducing power consumption without sacrificing quality-of-service. To solve the relaxed problem efficiently, we use a three-operator splitting method. Unlike benchmark methods that assume full antenna usage, our method makes fine-grained antenna-level energy decisions. This allows our algorithm to maintain low BER performance while significantly reducing the number of active RF chains, leading to substantial energy savings.

\vspace{-1.5mm}
\section{System Model}
\label{sec:sys_model}
\vspace{-1.5mm}

We consider the downlink of cell-free massive MIMO setup. There are $L$ APs and $K$ UEs that are randomly distributed in the coverage area. All UEs have a single antenna while each AP is equipped with $N$ antennas. We let $\vect{h}_{kl}\in \mathbb{C}^{N}$ denote the channel between UE $k$ and AP $l$. In this work, we focus on perfect channel state information (CSI) at the APs to focus solely on the adverse impact of the one-bit DACs. Concatenating all the channel vectors into a matrix, we obtain 
\begin{align}
    \vect{H} = \begin{bmatrix} \vect{h}_{11}^{\Ttran} & \cdots & \vect{h}_{1L}^{\Ttran} \\ \vdots & \ddots & \vdots \\ \vect{h}_{K1}^{\Ttran} & \cdots & \vect{h}_{KL}^{\Ttran} \end{bmatrix} \in \mathbb{C}^{K\times M}
\end{align}
where $M\triangleq LN$. We let $\vect{x}$ denote the quantized precoding signal with one-bit DACs. Assuming 1-bit DACs, the entries of $\vect{x}$ are taken as $\pm \sqrt{\frac{P}{2N}}\pm\imagunit\sqrt{\frac{P}{2N}}$, where $P$ is the total transmit power from an AP. In this paper, we also allow the entries of $\vect{x}$ to take the value of zero corresponding to the shutdown of the corresponding antenna element. The received signal at the UEs is given as
\begin{align}
\vect{y} = \vect{H}\vect{x}+\vect{n}
\end{align}
where $\vect{n}\sim \CN(\vect{0},\sigma^2\vect{I}_K)$ is the additive noise.   We denote the $k$-th entry of $\vect{y}$ by $y_k$ and assume that each receiver multiplies its received signal by a positive scaling factor $\beta>0$ to obtain the soft estimate of its data as
\begin{align}
    \hat{s}_k = \beta y_k
\end{align}
where $s_k$ is the complex symbol aimed for UE $k$ and $\hat{s}_k$ is the corresponding soft estimate. We define $\vect{s}=[s_1 \ \cdots \ s_K]^{\Ttran}$.
 Our aim is to design the 1-bit quantized precoding vector $\vect{x}$ so that we minimize the distance
\begin{align}
   \mathbb{E} \left\{\left\Vert \beta\vect{y}-\vect{s}\right\Vert^2\right\} \label{eq:expected-error}.
\end{align}
 The expectation in \eqref{eq:expected-error} with respect to the random noise realizations can be expressed as
\begin{align}
    \left \Vert  \vect{s}-\beta\vect{H}\vect{x}\right\Vert^2+\sigma^2 K\beta^2.
\end{align}

\section{Green One-Bit Quantization}

We define $\vect{b}=\beta\vect{x}$ as the optimization variable and define the set $\mathcal{B}=\left\{0,\pm \beta \sqrt{\frac{P}{2N}}\pm \imagunit\beta \sqrt{\frac{P}{2N}}\right\}$. In this work, we aim to minimize the energy consumption of the network, without sacrificing the quality-of-service of the UEs. This can be formulated as  the following multi-objective optimization problem:
\begin{align}
    \underset{\vect{b}\in \mathcal{B}^M}{\textrm{minimize}} \quad  \left \Vert  \vect{s}-\vect{H}\vect{b}\right\Vert^2 + \frac{NK\sigma^2 }{P}\left \Vert \vect{b} \right\Vert_{\infty}^2 + \lambda \|\vect{b}\|_0 \label{eq:problem1}
\end{align}
where we used the fact that at least one antenna is active, and for a $\vect{b}\in \mathcal{B}^M$, we have $\left\Vert \vect{b} \right\Vert_{\infty}^2=\beta^2\frac{P}{N}$. $\lambda \|\vect{b}\|_0 $ is the regularization term, aiming to deactivate the antennas, where $\lambda>0$ determines the trade-off between energy saving and performance of UEs. This formulation treats all UEs equally in terms of distortion minimization; however, depending on the structure of the channel matrix $\vect{H}$ and the input symbol vector $\vect{s}$, some UEs may be better served due to more favorable channel conditions. Additionally, as the number of UEs increases, the optimization becomes more constrained, making it harder to meet all UEs' quality-of-service targets while saving energy—thereby requiring careful tuning of $\lambda$ to balance performance and energy savings. In numerical studies, we prioritized the quality-of-service of the UEs and set the penalty coefficient, $\lambda$, accordingly. We now express the problem in terms of real-valued vectors and matrices by defining
\begin{align}
    \vect{s_r} = \begin{bmatrix} \Re(\vect{s}) \\ \Im(\vect{s}) \end{bmatrix}, \quad \vect{b_r}=\begin{bmatrix} \Re(\vect{b}) \\ \Im(\vect{b}) \end{bmatrix} 
\end{align} 
and
\begin{align}
    \vect{H_r}= \begin{bmatrix} \Re(\vect{H}) & -\Im(\vect{H}) \\
    \Im(\vect{H}) & \Re(\vect{H}) \end{bmatrix}.
\end{align}
We also define $\mathcal{B}_r=\left\{0,\pm \beta \sqrt{\frac{P}{2N}}\right\}$. Then the problem in \eqref{eq:problem1} can be expressed as
\begin{align}
    \underset{\vect{b_r}\in \mathcal{B}_r^{2M}}{\textrm{minimize}} \quad  \left \Vert  \vect{s_r}-\vect{H_r}\vect{b_r}\right\Vert^2 + \frac{2NK\sigma^2 }{P}\left \Vert \vect{b_r} \right\Vert_{\infty}^2 + \lambda \|\vect{b}\|_0. \label{eq:problem2}
\end{align}
The problem in \eqref{eq:problem2} is non-convex and combinatorial due to the limited quantization levels of the precoding vector. To encourage the deactivation of unnecessary antennas and convexify the problem,  we incorporate group lasso regularization, also known as sum-of-norms regularization \cite{parikh2014proximal}. This is achieved by summing the $l_2$-norms of the vectors formed by the real and imaginary components of each antenna. In this way, all elements within the vectors are aimed to be pushed to zero, while the sparsity among vectors is obtained. To implement this, we define 
 \begin{align}
     \overline{\vect{b}}_m = \left[\left[\vect{b_r}\right]_{m}, \ \left[\vect{b_r}\right]_{M+m}\right]^{\Ttran}\in \mathbb{R}^2.
 \end{align}
 The modified new problem becomes
\begin{align}
    \underset{\vect{b_r}\in \mathcal{B}_r^{2M}}{\textrm{minimize}} \, \left \Vert  \vect{s_r}-\vect{H_r}\vect{b_r}\right\Vert^2 + \frac{2NK\sigma^2 }{P}\left \Vert \vect{b_r} \right\Vert_{\infty}^2+\lambda\sum_{m=1}^M\left\Vert\overline{\vect{b}}_m\right\Vert_2.\label{eq:problem3}
\end{align}
To solve this problem efficiently, we drop the constraints $\vect{b_r}\in \mathcal{B}_r^{2M}$. The relaxed problem is
\begin{align}
    \underset{\vect{b_r}\in \mathbb{R}^{2M}}{\textrm{minimize}} \,  \left \Vert  \vect{s_r}-\vect{H_r}\vect{b_r}\right\Vert^2 + \frac{2NK\sigma^2 }{P}\left \Vert \vect{b_r} \right\Vert_{\infty}^2+\lambda\sum_{m=1}^M\left\Vert\overline{\vect{b}}_m\right\Vert_2\label{eq:problem4}.
\end{align}
The problem in \eqref{eq:problem4} is convex and can be solved using numerical solvers. To increase the computational efficiency further, we will resort to Forward Douglas-Rachford splitting, which is also known as Davis-Yin splitting \cite{davis2017three}. This algorithm can be applied since the first part of the objective function, which is $\left \Vert  \vect{s_r}-\vect{H_r}\vect{b_r}\right\Vert^2$ has a
Lipschitz-continuous gradient with Lipschitz constant $2\sigma^2_{\rm max}(\vect{H_r})$, i.e., two times the square of the largest singular value of $\vect{H_r}$. Moreover, the other two terms are proper closed convex functions with computable proximal operators. We split the objective function into three parts as
\begin{align}
 &   d_1(\vect{b_r}) = \left \Vert  \vect{s_r}-\vect{H_r}\vect{b_r}\right\Vert^2, \\
 & d_2(\vect{b_r})=\frac{2NK\sigma^2 }{P}\left \Vert \vect{b_r} \right\Vert_{\infty}^2, \\
 & d_3(\vect{b_r}) =\lambda\sum_{m=1}^M\left\Vert\overline{\vect{b}}_m\right\Vert_2. 
\end{align}
The method iteratively applies the proximal operators of $d_2$ and $d_3$ and uses gradient steps on $d_1$ to update $\vect{b}_r$. The update rules in (16)–(18) follow directly from this decomposition at the $i$th iteration \cite{anshika2024three}: equation (16) computes a proximal step for the group sparsity term $d_3$, (17) computes a proximal step for $d_2$ with a gradient descent correction for $d_1$, and (18) updates the dual variable as part of the algorithm's feedback mechanism. This formulation ensures efficient convergence while handling both smooth and nonsmooth terms in the objective. 
\begin{align}
 &   \vect{a_r}^{(i)} = \textrm{prox}_{\gamma d_3}\left(\vect{c_r}^{(i)}\right) \\
 & \vect{b_r}^{(i)} = \textrm{prox}_{\gamma d_2}\left(2\vect{a_r}^{(i)}-\vect{c_r}^{(i)}-\gamma\nabla_{d_1}\left(\vect{a_r}^{(i)}\right)\right) \\
 & \vect{c_r}^{(i+1)} = \vect{c_r}^{(i)}+ \psi(\vect{b_r}^{(i)}-\vect{a_r}^{(i)})
\end{align}
where the gradient $\nabla_{d_2}\left(\vect{a_r}^{(i)}\right)$ is given as
\begin{align}
 \nabla_{d_1}\left(\vect{a_r}^{(i)}\right)= 2\vect{H_r}^{\Ttran}\left(\vect{H_r}\vect{\vect{a_r}^{(i)}}- \vect{s_r}\right).  
\end{align}
The proximal operator $\textrm{prox}_{d_3}\left(\vect{c_r}^{(i)}\right)$ is given from \cite{parikh2014proximal} as
\begin{align}
&\left[\textrm{prox}_{\gamma d_3}\left(\vect{c_r}^{(i)}\right)\right]_m \nonumber\\
&= \max\left(0,1-\frac{\gamma\lambda}{\left \Vert \left [\left[\vect{c_r}^{(i)}\right]_m \  \left[\vect{c_r}^{(i)}\right]_{M+m} \right]^{\Ttran}\right\Vert_2}\right)\left[\vect{c_r}^{(i)}\right]_m, \\
&\left[\textrm{prox}_{\gamma d_3}\left(\vect{c_r}^{(i)}\right)\right]_{M+m} \nonumber\\
&= \max\left(0,1-\frac{\gamma\lambda}{\left \Vert \left [\left[\vect{c_r}^{(i)}\right]_m \  \left[\vect{c_r}^{(i)}\right]_{M+m} \right]^{\Ttran}\right\Vert_2}\right)\left[\vect{c_r}^{(i)}\right]_{M+m},
\end{align}
for $m=1,\ldots,M$. The proximal operator $\textrm{prox}_{\gamma d_2}\left(2\vect{a_r}^{(i)}-\vect{c_r}^{(i)}-\gamma\nabla_{d_1}\left(\vect{a_r}^{(i)}\right)\right)$ is computed from \cite{Jacobsson2017a} as in Algorithm~\ref{alg:1}.

\begin{algorithm}[t]
\caption{Proximal operator for the $l_{\infty}^2$-norm}\label{alg:1}
\begin{algorithmic}[1]
\State \textbf{inputs}: {$2\vect{a_r}^{(i)}-\vect{c_r}^{(i)}-\gamma\nabla_{d_1}\left(\vect{a_r}^{(i)}\right) \in \mathbb{R}^{2M},$ }
\State $ \vect{e} \gets \mathrm{abs}(2\vect{a_r}^{(i)}-\vect{c_r}^{(i)}-\gamma\nabla_{d_1}\left(\vect{a_r}^{(i)}\right))$ 
\State $ \vect{f} \gets \mathrm{sort}(\vect{e}, $`descending'$)$  
\For{$m = 1,\ldots,2M$}
\State $g_m \gets \frac{1}{\frac{4NK\sigma^2\gamma}{P} +m}\sum_{j=1}^m f_j $
\EndFor
\State $\alpha \gets\max\big(0, \max_m g_m\big)$
\For{$m = 1,\ldots,2M$}
\State $u_m \gets \min\left(e_m,\alpha\right) 
\cdot\mathrm{sgn}\left(\left[2\vect{a_r}^{(i)}-\vect{c_r}^{(i)}-\gamma\nabla_{d_1\gamma}\left(\vect{a_r}^{(i)}\right)\right]_m\right)$
\EndFor
\State \textbf{return} $\vect{u}$
\end{algorithmic}
\end{algorithm}

\section{Numerical Experiments and Results}
\label{sec:num_experiment}
In numerical analysis, a downlink cell-free communication setup is implemented in MATLAB. The simulation parameters for the system and the proposed algorithm are provided in Table~\ref{tab:sim_parameters}. Correlated Rayleigh fading model is considered for all the channels with spatial correlation matrices generated according to the local scattering model \cite{massivemimobook}. For each value of the regularization parameter $\lambda$, the average BER for each UE is computed over 10 different channel realizations, along with the corresponding number of active antennas. In each simulation instance, a unique channel realization is generated, and a total of $10^6$ uncoded data bits are transmitted for each UE using QPSK modulation. This corresponds to $5 \times 10^5$ complex symbols per UE. The BER for each UE is computed by counting erroneous bits and averaging over these $10^6$ bits per realization. After repeating this process for 10 different channel realizations, the final average BER for each UE is obtained by averaging across these 10 values. Note that in the performance figures, UEs are ranked based on their average BER values across all realizations to provide a clearer view of performance distribution among UEs.

 As benchmarks, 1-bit quantized versions of SQUID and regularized zero forcing (RZF) precoders (which is 1RZF) are implemented. Moreover, to enable a fair comparison under an antenna selection scenario, the antennas deactivated by our proposed algorithm are also deactivated in SQUID and RZF. ZF was excluded from this comparison because setting the corresponding columns of the channel matrix to zero leads to ill-conditioned matrices.

Additionally, we include a benchmark named ACR (antenna closed randomly), in which antennas are randomly deactivated until the same number of active antennas as in our algorithm is reached. It is worth noting that SQUID was originally developed for cellular multi-user massive MIMO systems, where it generally outperforms linear precoding techniques. However, in this study, SQUID was applied to a cell-free multi-user massive MIMO scenario, so deviations in its performance are to be expected. Perfect CSI is assumed throughout the simulations, and channel realizations were generated following the methodology described in \cite{SIG-109}.

\begin{figure}[t!]
		\begin{center}
			\includegraphics[trim={8mm 2mm 10mm 8mm},clip,width=3.6in]{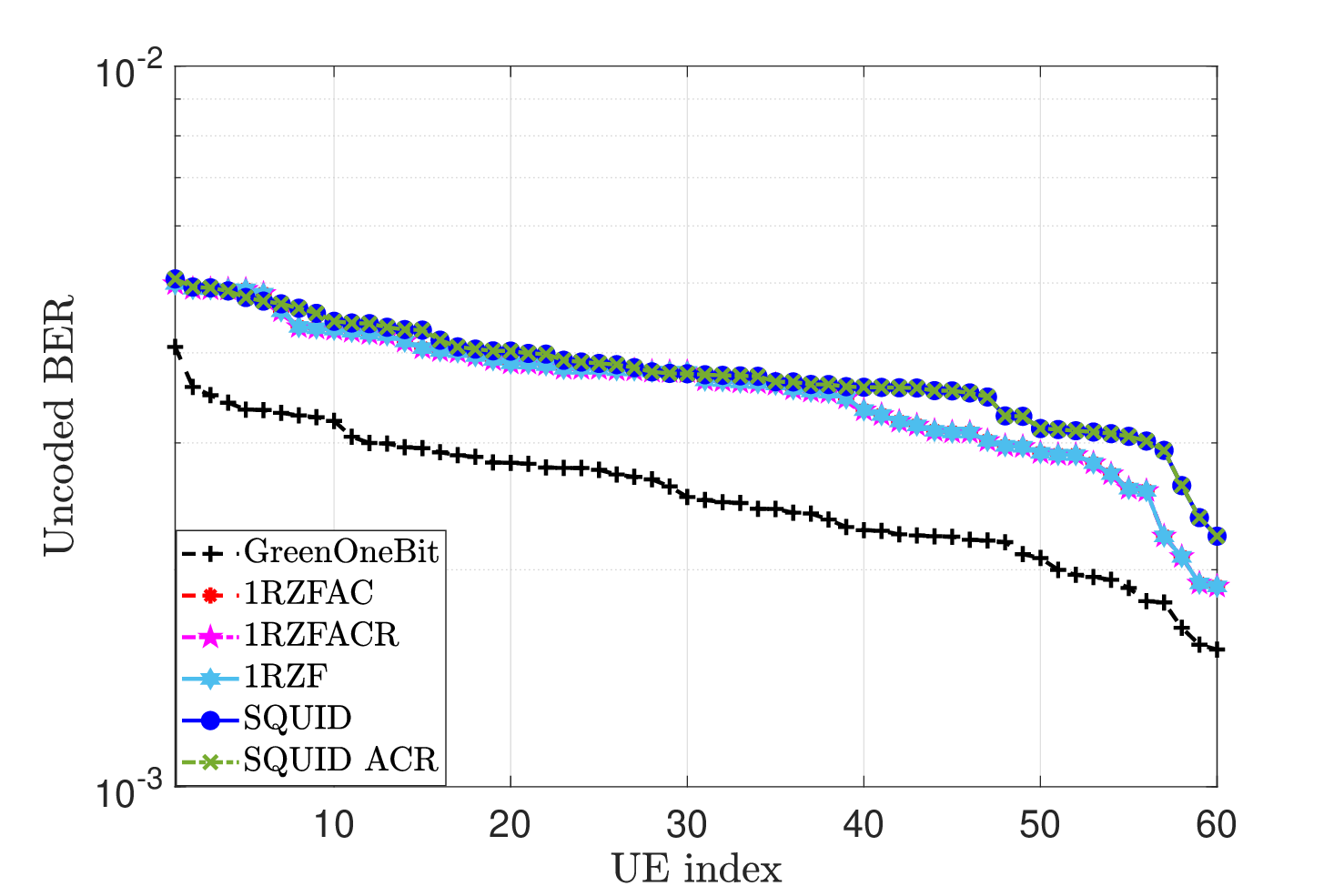}
			\caption{Uncoded BER performance of the UEs for $\lambda = 1$.} \label{fig:fig1}
		\end{center}
\end{figure}

Fig.~\ref{fig:fig1} presents results for $\lambda = 1$. This corresponds to a scenario where UE performance is prioritized over energy savings by antenna deactivation. As expected, this results in nearly all antennas being active (as can also be verified by Fig.~\ref{fig:fig4}). Consequently, the BER performance of the proposed algorithm is superior to the other benchmark methods despite minimal antenna deactivation.  This highlights an important property of the proposed algorithm: even when antenna deactivation is minimal, it can still outperform other methods in terms of BER. It is worth noting that the BER performance of RZF and RZF-ACR are nearly identical. This is because in our simulation setup, RZF-ACR is configured to use exactly the same number of active antennas as our proposed algorithm, but selects them randomly. As a result, both RZF and RZF-ACR operate under nearly identical antenna budgets in these scenarios.

\begin{figure}[t!]
		\begin{center}
			\includegraphics[trim={8mm 2mm 10mm 8mm},clip,width=3.6in,height=2.5in]{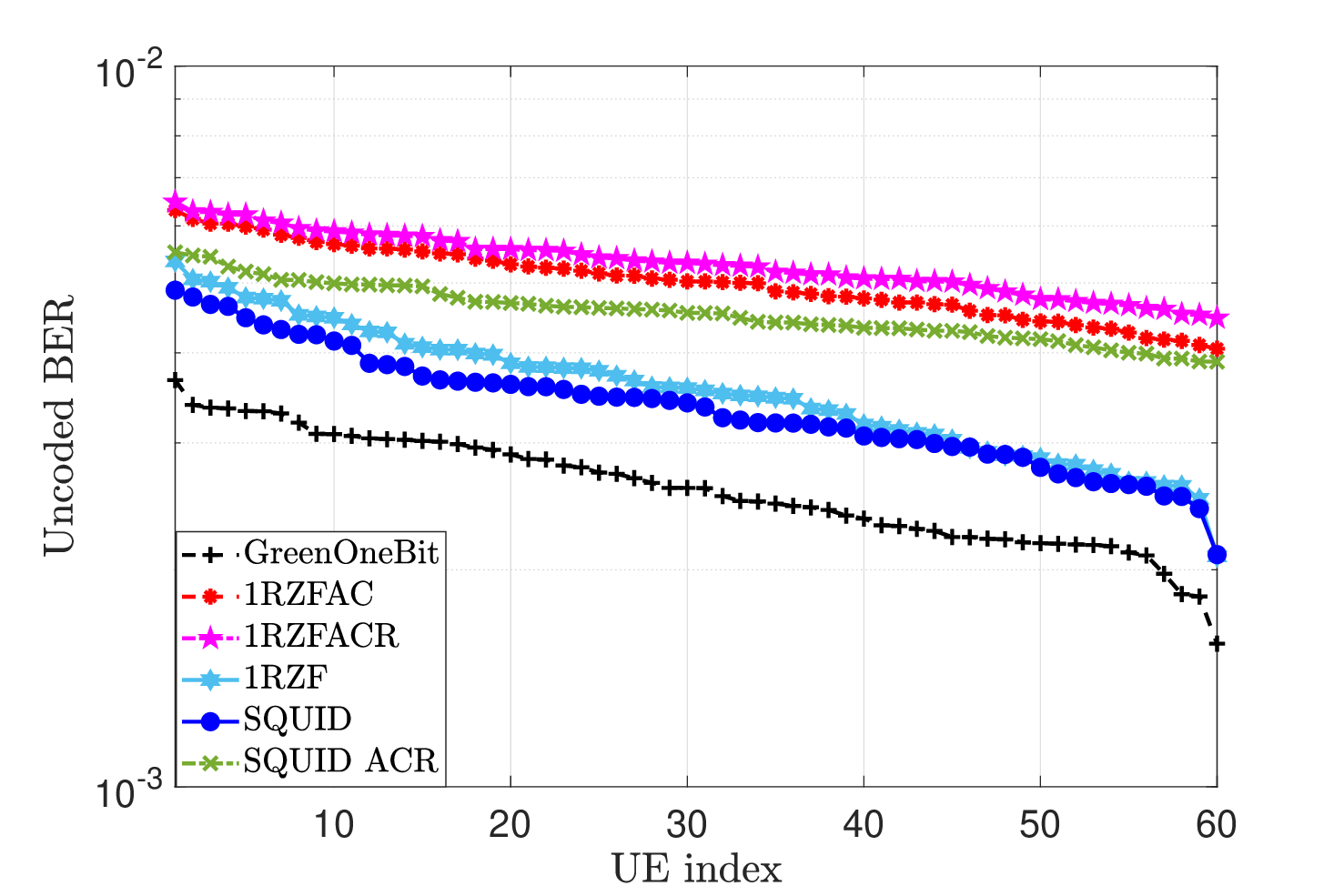}
			\caption{Uncoded BER performance of the UEs for $\lambda = 15$.} \label{fig:fig2}
		\end{center}
\end{figure}

Fig.~\ref{fig:fig2} shows the BER for $\lambda = 15$. When compared with the corresponding antenna activity in Fig.~\ref{fig:fig4}, it is evident that the proposed algorithm maintains high performance even while deactivating an average of 20 out of 100 antennas. Unlike Fig.~\ref{fig:fig1}, the BER of the other algorithms degrades due to the antenna deactivation. Notably, random antenna deactivation (ACR) leads to significantly worse BER, whereas the proposed method yields superior performance. This demonstrates the effectiveness of our algorithm in identifying which antennas to deactivate.

\begin{figure}[t!]
		\begin{center}
			\includegraphics[trim={8mm 2mm 10mm 8mm},clip,width=3.6in,height=2.5in]{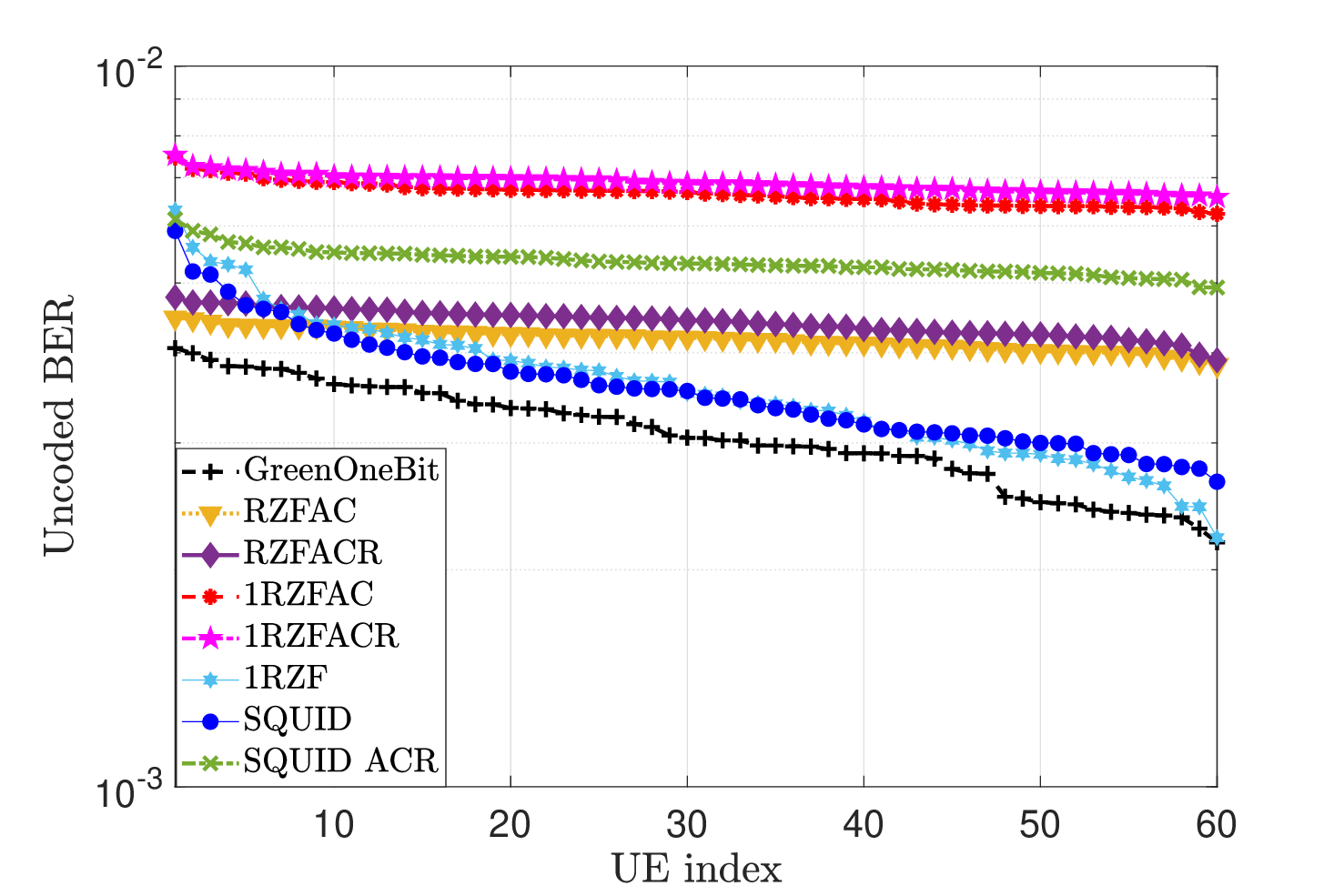}
			\caption{ Uncoded BER performance of the UEs for $\lambda = 25$.} \label{fig:fig3}
		\end{center}
\end{figure}

\begin{figure}[t!]
		\begin{center}
			\includegraphics[trim={8mm 2mm 10mm 8mm},clip,width=3.6in,height=2.5in]{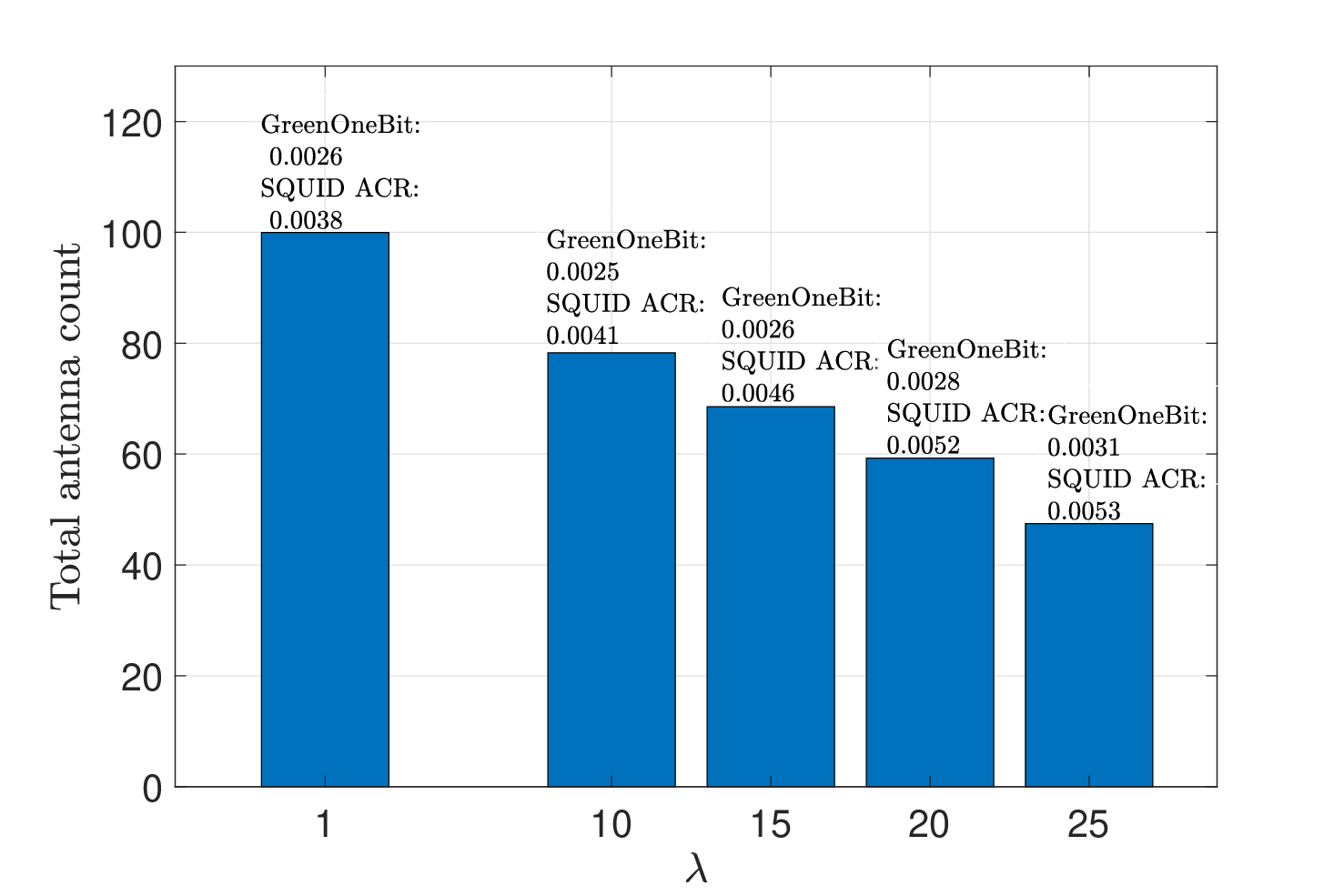}
			\caption{Average active antenna count and uncoded BER.} \label{fig:fig4}
		\end{center}
\end{figure}

In the case of $\lambda = 25$, shown in Figure~\ref{fig:fig3}, the number of active antennas decreases further. Despite this, the proposed algorithm sustains its performance, while the baseline methods experience a more pronounced degradation in BER. At this point, nearly half of the antennas are inactive, which underscores the energy efficiency potential of our approach. This behavior arises because the group sparsity-promoting regularization is applied to the optimization variable $\vect{b}$, which encodes the scaled and quantized transmitted signal. Since $\vect{b}$ is determined anew for each transmitted symbol vector $\vect{s}$, the optimization adaptively selects which antenna signals (and thus APs) should be active based on the structure of that symbol vector. Antennas that contribute minimally to reconstructing $\vect{s}$ under the channel realization $\vect{H}$ tend to be deactivated to reduce energy usage. Therefore, even though the AP selection is not explicitly hard-coded, the solution to the optimization problem results in a dynamic and data-dependent antenna activation pattern—effectively tailoring the active AP set to each symbol vector.
Figs.~\ref{fig:fig2} and \ref{fig:fig3} also show that RZF performs the worst under antenna deactivation, highlighting its sensitivity to changes in the active antenna set.

Finally, Fig.~\ref{fig:fig4} illustrates the average number of active antennas for each $\lambda$ value. This figure demonstrates that the proposed algorithm can significantly reduce power consumption while maintaining similar BER performance. For example, at $\lambda = 25$, approximately 50\% of the total antenna power is saved without a significant loss in communication quality.
\renewcommand{\arraystretch}{1.3} 
\begin{table}[h]
\centering
\begin{tabular}{|c|c|}
\hline
Parameter & Value(s) \\
\hline
$\lambda$ & 1, 10, 15, 20, 25\\
\hline
Network area & 1 km x 1km\\
\hline
Network layout & uniform AP distribution\\
& (with wrap around)\\
\hline
Number of APs, $L$ & 100\\
\hline
Number of antennas per AP, $N$  & 1\\
\hline
Number of UEs, $K$ & 60\\
\hline
Bandwidth & 20 MHz\\
\hline
Height diff. between APs and UEs & 10 m\\
\hline
Azimuth angular standard deviation  & 15$^\circ$\\
\hline
Elevation angular standard deviation  & 15$^\circ$\\
\hline 
Channel gain at 1\,km & -140.6 dB\\
\hline
Pathloss exponent & 3.67\\
\hline
Standard deviation of shadow fading & 4\\
\hline
Downlink power & 1\,W\\
\hline
Number of bits per UE & $10^6$\\
\hline
Number of setups & 10\\
\hline
$\gamma$ & $\frac{0.1}{2\sigma^2_{\rm max}(\vect{H_r})}$\\
\hline
Modulation type & QPSK\\
\hline
\end{tabular}
\caption{Simulation Parameters}
\label{tab:sim_parameters}
\end{table}

 \section{Conclusion}
In this work, we propose a green 1-bit quantized precoding algorithm for cell-free massive MIMO networks. The proposed algorithm jointly minimizes the MSE of the UEs and the number of active antennas in the system. We first used group-sparsity based convex relaxation to the original non-convex combinatorial multi-objective problem. To increase the computation efficiency further,  we used a three-operator splitting method based on SQUID algorithm. Our results have shown that the proposed algorithm effectively reduces both the BER of the UEs and the number of active antennas in the system. Compared to the benchmark schemes, the proposed algorithm can reduce the BER by 25\%, while effectively reducing the energy consumption by 50\%.  As future work, the cost of antenna activation/deactivation can be included in the dynamic arriving/departing UE model, where longer-term antenna deactivation methods can be proposed to reduce long-term energy consumption. Moreover, while the proposed algorithm converges approximately 10 times faster than the CVX solution, it converges approximately 40 times slower than SQUID. Iteration count is affecting the convergence of the algorithm. When the count is small, BER increases. Increasing the iteration count dramatically decreases the BER while the convergence time increases. A computationally more efficient algorithm can be investigated in the future.

\bibliographystyle{IEEEtran}

\bibliography{IEEEabrv,refs}

\end{document}